\documentclass[showpacs,superscriptaddress,preprintnumbers,amsmath,amssymb,onecolumn,nofootinbib]{revtex4}
\usepackage{graphicx}% Include figure files
\usepackage{dcolumn}% Align table columns on decimal point
\usepackage{bm}% bold math
\usepackage{type1cm}

\newcommand{\varphim}{\!{\stackrel{\,_{(m)}}\varphi}\!}

\newcommand{\varphizero}{\!{\stackrel{\,_{(0)}}\varphi}\!}

\numberwithin{equation}{section}
\begin{document}

\preprint{YITP-08-81}

\title{Efficient diagrammatic computation method for higher order correlation functions of local type primordial curvature perturbations}% Force line breaks with \\

\author{Shuichiro Yokoyama}
\email{shu@a.phys.nagoya-u.ac.jp}
\affiliation{Department of Physics and Astrophysics, Nagoya University, Aichi 464-8602, Japan}%Lines break automatically or can be forced with \\  

\author{Teruaki Suyama}%
\email{suyama@icrr.u-tokyo.ac.jp}
\affiliation{Theoretical and Mathematical Physics Group,
Center for Particle Physics and Phenomenology,
Louvain University,
2 Chemin du Cyclotron, 1348 Louvain-la-Neuve, Belgium}

\author{Takahiro Tanaka}
\email{tama@scphys.kyoto-u.ac.jp}
\affiliation{Yukawa Institute for Theoretical Physics, Kyoto University, Kyoto 606-8502, Japan} 
\date{\today}% It is always \today, today,
             %  but any date may be explicitly specified

\begin{abstract}
%Using diagrammatic method,
We present a new efficient method for computing the non-linearity parameters of the
higher order correlation functions of local type curvature perturbations in inflation models having a $\cal N$-component scalar field, focusing on the non-Gaussianity generated during the evolution on super-horizon scales. 
In contrast to the naive expectation that the number of operations necessary to compute the $n$-point functions is proportional to ${\cal N}^n$, it grows only linearly in ${\cal N}$ in our formalism. Hence, our formalism is particularly powerful for the inflation models composed of a multi-component scalar field, including the models in which
the slow-roll conditions are violated after the horizon crossing time. 
Explicit formulas obtained by applying our method are provided for $n=2,3,4$ and 5, which correspond to  power-, bi-, tri- and {\it quad}-spectra, respectively. We also discuss how many parameters we need to parameterize the amplitude and the shape
of the higher order correlation functions of local type.
\end{abstract}

% PACS, the Physics and Astronomy
                             % Classification Scheme.
%\keywords{Suggested keywords}%Use showkeys class option if keyword
                              %display desired
\maketitle

\section{Introduction}
\label{intro}
Current observations of the cosmic microwave background (CMB) anisotropies indicate that
primordial curvature perturbations are almost Gaussian~\cite{Komatsu:2008hk}.
In general, if the perturbations are purely Gaussian, 
the statistical properties of the perturbations can be completely described by the two-point 
correlation function (=power spectrum).
On the other hand, 
if the perturbations deviate from the Gaussian distribution, 
the non-Gaussianity affects the higher order correlation functions, or higher order spectra.
Currently, 
the non-Gaussianity is attracting attention as a powerful probe to discriminate various inflation models~\cite{Komatsu:2001rj,Bartolo:2004if}.
In particular, there are a large number of studies on the three-point
correlation function 
(=bi-spectrum).
However, the four-point correlation function (=tri-spectrum)
can also be constrained by future accurate measurements~\cite{Okamoto:2002ik,Bartolo:2005fp,Kogo:2006kh,D'Amico:2007iw}.
Using the analysis of both the bi-spectrum and the tri-spectrum 
in the future experiments, it is expected that
we can extract more information about the mechanism of generating the primordial curvature perturbations.
Hence, it is important to obtain useful formulas for the higher order
correlation functions of primordial curvature perturbations.%~\cite{Byrnes:2007tm}.

Roughly speaking, the leading order of the connected part of $n$-point function
is $O(P^{n-1})$, where $P \sim 10^{-10} $ is the amplitude of the power
spectrum. Hence, it is naively
expected to be difficult to measure higher order correlation functions. 
However, when the non-Gaussianity is large,
this estimate $O(P^{n-1})$ will be replaced with $O(f_{{\rm NL}}^{n-2} P^{n-1})$
or even larger.
Here,
$f_{{\rm NL}}$ is a non-linearity parameter given in Refs.~\cite{Komatsu:2001rj,Bartolo:2004if}; 
%$f_{{\rm NL}}$ is a non-linearity parameter which has been originally introduced by Ref.~\cite{Komatsu:2001rj} as
\begin{eqnarray}
\zeta = \zeta_G + {3 \over 5}f_{{\rm NL}} \zeta_G^2~,
\label{NLoriginal}
\end{eqnarray}
where $\zeta$ is the curvature perturbation on uniform energy density hypersurface and $\zeta_G$ is the linear Gaussian part. 
Notice that observationally $f_{{\rm NL}}$ can be as large as $O(100)$.
This possible enhancement slightly improves the detectability
of the higher order correlation functions. Furthermore, the
number of argument wavenumbers of the $n$-point function is $n-1$. 
When the CMB temperature anisotropies, $C_\ell$, are measurable up to
$\ell=\ell_{\rm max}$, the number of independent wavenumbers which we can measure will
be roughly estimated as $\ell_{\rm max}^2$. Hence, the number of different
combinations of argument wavenumbers increases as $\ell_{\rm max}^{2(n-1)}$.
This large number enhances the effective amplitude of $n$-point function to
$O(f_{{\rm NL}}^{-1} (f_{{\rm NL}}\ell_{max} P)^{n-1})$, while the amplitude of Gaussian
noise is $O(P^{n/2})$. The detectability of the $n$-point function is
basically determined by the ratio of these two numbers,
$O\left(f_{{\rm NL}}^{-1} P^{-1/2} (f_{{\rm NL}}\ell_{max} \sqrt{P})^{n-1}\right)$.
Hence, if $f_{{\rm NL}}\ell_{max} \sqrt{P}$ exceeds unity, all the higher
order correlation functions are in principle measurable.
For the Planck satellite~\cite{:2006uk}, it is expected that $\ell_{max} \sim O(2000)$.
Hence, naively, if $f_{{\rm NL}}$ would be as large as $O(50)$,
$f_{{\rm NL}} \ell_{max} \sqrt{P}$ can exceed unity.
This fact strongly motivates a systematic derivation of the formulas for
higher order correlation functions.

In this paper, 
we present a new method to calculate general $n$-point functions of
local type primordial curvature perturbations. 
This new method is much more efficient than the straightforward
calculations, especially when applied to the models with many components of inflaton field, including the models in which
the slow-roll conditions are violated after the horizon crossing time.
This method is based on the diagrammatic approach given in Ref.~\cite{Byrnes:2007tm} 
as well as on our previous work~\cite{Yokoyama:2007uu,Yokoyama:2007dw}, 
in which the formulation for the bi-spectrum was developed.
As for the parameterization of the higher order spectra, it is well known that
the bi-spectrum can be parameterized by a single parameter, 
so-called non-linearity parameter, $f_{NL}$,
while the tri-spectrum is parameterized by two parameters $\tau_{NL}$ and $g_{NL}$~\cite{Byrnes:2006vq} 
due to the existence of two distinct terms that exhibit a different wavenumber dependence.
That is, 
the number of parameters necessary to describe the higher order correlation functions
is equal to the number of independent terms which have a different wavenumber dependence.
Based on the diagrammatic method, 
we also show that one can easily count how many parameters we need to parameterize the amplitude and the shape
of higher order spectra of local type.

This paper is organized as follows.
In section \ref{diagram} we briefly review 
the $\delta N$ formalism~\cite{Starobinsky:1986fx,Sasaki:1995aw,Sasaki:1998ug,Lyth:2004gb,Lyth:2005fi}, 
which is the foundation of our present analysis. 
We also discuss how many parameters we need to parameterize the higher
order correlation functions. 
In section \ref{diagram2} 
we present our diagrammatic method for the computation of 
$n$-point correlation functions of primordial curvature perturbations. 
As an application of our method, in the succeeding section \ref{examplesec}
we give concise formulas for the power-, bi-, tri- and {\it quad}-spectra
of the primordial curvature perturbations generated in multi-component 
inflation models. 
Section \ref{sum} is devoted to discussion and conclusion. 

\section{local type primordial curvature perturbations and their parameterization}
\label{diagram}

We focus on the non-Gaussianity generated 
during the evolution on super-horizon scales in multi-scalar inflation. 
We start with a brief review of the $\delta N$ formalism.
Using the $\delta N$ formalism, we present a diagrammatic representation %%changed
for general $n$-point functions of local type primordial curvature
perturbations, and show how they are parameterized.%~\cite{Byrnes:2007tm}.

\subsection{Background equations}

We consider a ${\cal N}$-component scalar field whose action is given by
\begin{eqnarray}
S = - \int d^4 x \sqrt{-g}\left[{1 \over 2}h_{IJ}g^{\mu\nu}\partial_{\mu}\phi^I\partial_{\nu}\phi^J + V(\phi)\right]~,&&\\
\left(I,J=1,2,\cdots,{\cal N}\right)~,&&\nonumber
\end{eqnarray}
where $g_{\mu\nu}$ is the spacetime metric and $h_{IJ}$ is the metric on the scalar field space. 
In this paper 
we restrict our discussion to the flat field space metric
$h_{IJ}=\delta_{IJ}$
%{\bf (canonical field)}%%changed 
to avoid inessential complexities due to non-flat field space metric, though
the generalization is straightforward~\cite{Yokoyama:2007dw}.

We define $\varphi^I_{i} (i=1,2)$
as\footnote{
Here, we take different definition for $\varphi^I_2$ from that
introduced %%changed
in
our previous papers~\cite{Yokoyama:2007uu,Yokoyama:2007dw},
which was defined as $\varphi^I_2 \equiv d\phi^I/dN$.
Based on previous definition, specific expressions for $P^a_{~b}$ or $Q^a_{(\ell)b_1 b_2\cdots b_{\ell-1}}$
defined as Eq.~(\ref{PQ}) in the later Sec.~\ref{treeshaped}
include the terms which diverge when $V=0$,
which is not a suitable formulation for the numerical calculations.
If we define $\varphi^I_2$ as in Eq.~(\ref{phasevar}),
there are no divergences of $P^a_{~b}$ or $Q^a_{(\ell)b_{1}b_{2}\cdots
b_{\ell-1}}$ at the time when $V=0$. %%changed
}
\begin{eqnarray}
\varphi^I_1 \equiv \phi^I~,~~\varphi^I_2 \equiv \dot{\phi}^I~,
\label{phasevar}
\end{eqnarray}
where a dot `$`~\dot{}~$'' represents differentiation %%changed
with respect to the cosmological time.
%Hence, the explicit forms of $H_a$ and $H_{ab}$ are also different from
%the form given in~\cite{Yokoyama:2007dw}.
%{\it Why do you want to say this here? 
%Notice that, from these explicit expressions, one find that the local quantities $N_{a}^{F}$,
%$N_{ab}^{F}$ and $N_{abc}^{F}$ do not diverge even when $V=0$.

For brevity, hereinafter, we use Latin indices at the beginning of Latin
alphabet, $a$, $b$ or $c$, instead of the double indices, i.e., $X^a =X^I_i$.
Then, the background equation of motion for $\varphi^a$ is
\begin{eqnarray}
{d \over dN}\varphi^a = F^a(\varphi)~, \label{back1}
\end{eqnarray}
where $N$ is the $e$-folding number and
$F^a (= F^I_i)$ is given by  
\begin{eqnarray}
F^I_1 = {\varphi^I_2 \over H}~,
~~F^I_2 = -3\varphi^I_2 - {V^I \over H} ~, 
\end{eqnarray}
with $V^I=\delta^{IJ}(\partial V/\partial\phi^J)$. 
The homogeneous background Friedmann equation is given by 
\begin{equation}
H^2 = {1 \over 3}\left({1 \over 2}\varphi^I_2\varphi_{2I} + V\right)~, \label{friedmann}
\end{equation}
with $\varphi_{2I} = \delta_{IJ}\varphi^J_2$.

In the $\delta N$ formalism~\cite{Starobinsky:1986fx,Sasaki:1995aw,Sasaki:1998ug,Lyth:2004gb,Lyth:2005fi},
the difference in $e$-folding number %%changed 
between two adjacent background solutions describes the
evolution of $\zeta$, curvature perturbations, %%changed
on super-horizon scales.
The solution of the background inflationary dynamics dominated by a ${\cal N}$-component scalar field
is labelled by $2{\cal N}-1$ integration constants $\lambda^a$, besides the trivial time translation $\delta N$.
Let us define $\delta \varphi^a$ as the perturbation,
\begin{eqnarray}
\delta \varphi^a (\lambda;N) \equiv \varphi^a( \lambda+\delta\lambda;N)-\varphi^a(\lambda;N)~, \label{per1}
\end{eqnarray}
where $\lambda$ is abbreviation of $\lambda^a$
and $\delta\lambda^a$ is a small quantity of $O(\delta)$.
%KOKO

2${\cal N}$ parameters $\{N,\lambda^a\}$ parameterize the initial values of fields.
There is an arbitrariness in choosing the integral constants,
i.e. a different choice of integration constants ${\bar \lambda^a}
\equiv f^a (\lambda)$ is equally good. 
Here we leave the choice of $\lambda^a$ unspecified 
since all the discussion in the paper is not affected by the choice. 

$\delta \varphi^a (N)$ defined by Eq.~(\ref{per1}) represent %%changed
perturbations of the scalar field on the
$N= {\rm constant}$ gauge~\cite{Sasaki:1998ug}.
In this case, 
$\zeta$ at each point in space depends on 
the fluctuations of the scalar field at the same spatial point,
and is given by 
\begin{eqnarray}
\label{higher}
&&\zeta (N_F,{\vec x}) = \sum{1 \over n!} N^*_{a_1a_{2\cdots}a_n}
\delta \varphi^{a_1}_*({\vec x})\delta \varphi^{a_2}_*({\vec x}) \cdots
\delta \varphi^{a_n}_*({\vec x})~, \label{zeta}\\
&&N^*_{a_1a_{2\cdots}a_n} 
\equiv {\partial^n N (N_F, \{\varphi^a\}) 
\over \partial \varphi^{a_1} \partial \varphi^{a_2} \cdots
\partial \varphi^{a_n}}\biggr|_{\varphi^a=\varphizero^a(N_*)}~%KOKO
.\nonumber
\end{eqnarray}
Here,
the values of scalar fields on 
the initial flat hypersurface, $\{ \varphi^a_* \}$, 
differ from place to place and characterize the initial perturbation. 
Since the e-folding number between the initial
flat hypersurface and the final uniform energy density hypersurface 
depends on $\{ \varphi^a\}$, 
as its argument we have used $\{ \varphi^a\}$ instead of the initial
time $N=N_*$.
We have decomposed the scalar field as $\varphi^a = \varphizero^a+\delta \varphi^a$ and 
Taylor expanded $\zeta$ in terms of %%changed
$\delta \varphi^a=O(\delta)$. The suffix 
$*$ represents the value evaluated at a certain time $N_*$ which is 
shortly after the horizon crossing time.  
The final hypersurface at $N=N_F$ is chosen to be an uniform energy
density surface. 
As is well known, $\zeta(N_F)$ is independent of the choice of $N_F$ 
as long as $N_F > N_c$, where $N_c$ is a certain time after the background trajectories have 
completely converged.  
According to the $\delta N$ formalism, 
the expansion coefficients $N^*_{a_1a_{2\cdots}a_n}$ are simply given by
the derivatives of $N(N_F,\{\varphi^a\})$, where 
$N(N_F,\{\varphi^a\})$ 
is the $e$-folding number spent during the evolution of %KOKO
the homogeneous universe, in phase space, from
the initial point $\{\varphi^a\}$
to the final uniform energy density surface.

\subsection{Parameterization of the $n$-point functions}
\label{highersec}

Let us begin with the two-point function.  
At the leading order in $\delta$,\footnote{
In Refs.~\cite{Byrnes:2007tm,Cogollo:2008bi},
the authors have also considered the one-loop corrections
which are the higher order in $\delta$.
In this paper we consider the only tree-level spectrum
and neglect the one-loop corrections.}

the two-point function of $\zeta$ can be written as
\begin{eqnarray}
\langle \zeta_{{\vec k}_1} \zeta_{{\vec k}_2}\rangle_c &\equiv& P_\zeta (k_1) \delta^{(3)}({\vec k}_1+{\vec k}_2) \nonumber \\
&=&N_aN_{b}{\cal A}^{ab}(\vec k_1)\delta^{(3)}({\vec k}_1+{\vec k}_2)~, 
\label{higher1}
\end{eqnarray}
where $\langle \cdots \rangle_c$ 
means the expectation value of the connected 
part of ``$\cdots$'', and 
we have abbreviated the suffix $*$. 
Here we have introduced the covariance matrix ${\cal A}^{ab}(\vec k)$ 
defined by ${\cal A}^{ab}(\vec k)\delta (\vec k+\vec k')
\equiv\langle \delta \varphi^a_{*\vec k}
\delta
\varphi^b_{*\vec k'} \rangle_c$.%%changed 

We assume that all the relevant components of the scalar field satisfy
the slow-roll conditions at least until $N=N_*$, in which our 
formalism works quite efficiently. Otherwise, correlation functions 
can not be parameterized by a small number of 
parameters. %%changed (simply added)
In this case, $\{\delta \varphi^a_*\}$ 
is approximated by a set of Gaussian random variables
 with the scale
invariant spectrum
\footnote{
As is well known, the deviation from the scale invariant spectrum can be
given by the slow-roll parameters at the horizon crossing time of the
corresponding scale,  
and $A^{IJ}_{12}$, $A^{IJ}_{21}$ and $A^{IJ}_{22}$ are also suppressed
by the same slow-roll parameters.
Since we know that the deviation from the scale invariance is observationally
small, it is natural to assume that the slow-roll conditions are well satisfied 
at around the horizon crossing time. 
Therefore, for simplicity, we evaluated $A^{ab}$ at the horizon crossing time
in the limit of vanishing slow-roll parameters. 
}
, 
and 
${\cal A}^{ab}$ is given by ${\cal A}^{ab}=A^{ab} P(k)$ with 
\begin{eqnarray}
A^{IJ}_{11} &\!\! =&\!\!\delta^{IJ}~,
\qquad A^{IJ}_{12}=A^{IJ}_{21}=A^{IJ}_{22}=0~, 
 \label{aij}
\qquad P(k) = {2\pi^2 \over k^3}\left({H_* \over 2\pi}\right)^2~.
\end{eqnarray}
%and $\epsilon^{IJ}=0$.
Strictly speaking, even in the slow-roll inflation, 
$\delta \varphi^a_*$ deviates from pure Gaussian perturbation due to %KOKO
the effect of interaction. 
 However, 
%in Refs.~\cite{Seery:2006vu,Seery:2006js,Jarnhus:2007ia} the authors calculated the magnitudes of
the non-Gaussianity of $\zeta$ caused by this deviation
is suppressed by the slow-roll parameters,
which is an undetectable level in the future experiments \cite{Seery:2006vu,Seery:2006js,Jarnhus:2007ia,Arroja:2008ga}. 
Hence, we neglect the non-Gaussianity of $\delta \phi^I_*$ here.
\footnote{ Note that
in the slow-roll inflation the non-Gaussianity of $\delta \varphi^a_*$ dominates the trispectrum of $\zeta$.
But it is too small to be detectable in the future experiments~\cite{Seery:2006vu,Seery:2006js,Jarnhus:2007ia,Arroja:2008ga}.}
In a similar fashion,
the three-point correlation function at the leading
order is given as~\cite{Lyth:2005fi}
\begin{eqnarray}
\langle \zeta_{{\vec k}_1} \zeta_{{\vec k}_2} \zeta_{{\vec k}_3}\rangle_c 
=\frac{N^aN^bN_{ab}}{ {(N_c N^c)}^2} \left(P_\zeta(k_1)P_\zeta(k_2) +
				      2~{\rm perms.} \right)
\delta^{(3)}({\vec k}_1+{\vec k}_2+{\vec k}_3)~, \label{higher2}
\end{eqnarray}
where $N^a\equiv A^{ab}N_b$. 
In deriving this equation,
we have used 
$\langle \delta \varphi^a_{*{\vec k}_1} \delta \varphi^b_{*{\vec k}_2} \delta \varphi^c_{*{\vec k}_3}\rangle_c = 0$. 
From this equation, we find that
$\langle \zeta_{{\vec k}_1} \zeta_{{\vec k}_2} \zeta_{{\vec k}_3}\rangle_c$
is $O(\delta^4)$. 
Since the wavenumber %%changed
dependence of the bi-spectrum is completely 
given by the products of the power spectrum,
the bi-spectrum is characterized by a single parameter 
$N^a N^b N_{ab}/{(N^c N_c)}^2$, which controls the overall amplitude. 
Following literatures,
we redefine the non-linearity parameter $f_{\rm NL}$ given in the introduction as~\cite{Lyth:2005fi}
\begin{eqnarray}
f_{\rm NL}=\frac{5}{6} \frac{N^a N^b N_{ab}}{{(N_c N^c)}^2}.
 \label{higher3}
\end{eqnarray}
If only one field contributes to the curvature perturbation,
$f_{\rm NL}$ defined by Eq.~(\ref{higher3}) is equivalent to Eq.~(\ref{NLoriginal}).
While Eq.~(\ref{NLoriginal}) is valid only when the single field dominates
the curvature perturbation,
Eq.~(\ref{higher3}) can be applied to larger classes of inflation models where
the curvature perturbations are sourced by multiple fields.

We can also write down the leading order four-point correlation function
(the tri-spectrum) 
as~\cite{Lyth:2005fi,Byrnes:2006vq}
\begin{eqnarray}
&& \langle \zeta_{{\vec k}_1} \zeta_{{\vec k}_2} \zeta_{{\vec k}_3}
 \zeta_{{\vec k}_4}\rangle_c 
= \bigg[ \frac{N^aN_{ab}N^{bc}N_c}{ {(N_d N^d)}^3} \left( P_\zeta (k_1)
						    P_\zeta (k_{12})
						    P_\zeta (k_4) +
						    11~{\rm perms.}
						   \right) \nonumber\\
&& \hspace{30mm}
+\frac{N^aN^bN^cN_{abc}}{ {(N_d N^d)}^3} \left( P_\zeta (k_1) P_\zeta
					  (k_{2}) P_\zeta (k_3) + 3~{\rm
					  perms.} \right)
\bigg]\delta^{(3)}({\vec k}_1+{\vec k}_2+{\vec k}_3+{\vec k}_4)~,
\label{II-14}
\end{eqnarray}
where $k_{ij} \equiv |{\vec k}_i+{\vec k}_j|$ and $N^{ab}\equiv
A^{ac}A^{bd}N_{cd}$.
We see that the four-point function is $O(\delta^6)$.
Unlike the bi-spectrum,
the tri-spectrum has two distinct terms that exhibit different
wavenumber dependence.
As a consequence,
we need two parameters to specify the tri-spectrum.
Following Ref.~\cite{Byrnes:2006vq},
we use the non-linearity parameters $\tau_{\rm NL}$ and $g_{\rm NL}$
defined by
\begin{eqnarray}
\tau_{\rm NL} =\frac{N^aN_{ab}N^{bc}N_c}{ {(N_d N^d)}^3},~~~~~~~~g_{\rm
 NL}={25 \over 54}\frac{N^aN^bN^cN_{abc}}{ {(N_d N^d)}^3}.
\label{higher4}
\end{eqnarray}
We can further proceed to higher order correlation functions.
The main issue that we address in the rest of this section is how many
parameters
are necessary to parameterize the local type $n$-point function. 
Of course, 
we can count the number of such parameters by directly calculating
the $n$-point function from Eq.~(\ref{higher}) as in the case of the
bi-spectrum
or the tri-spectrum. 
Although in principle there are no difficulties in such a direct
counting,
the actual computation becomes exponentially more cumbersome 
as we proceed to higher order. 
Here, instead of resorting to the direct computation, 
we use a diagrammatic method~\cite{Byrnes:2007tm}.

The leading order of the $n$-point function consists of terms
of $O(\delta^{2n-2})$, 
which are given by products of $(n-1)$ power spectra.
According to the diagrammatic method,
each of these leading terms has a corresponding connected diagram that consists of $n$ vertices and $(n-1)$ lines 
connecting two vertices.
Such a connected diagram should have a tree structure. 
Namely, there is always a unique path that connects any pair of vertices
in the
diagram. 
We refer to such diagrams as {\it reduced} tree diagrams, 
to distinguish them from the ({\it full}) tree diagrams 
that will be introduced later.

\begin{figure}[htbp]
  \begin{center}
    \includegraphics[keepaspectratio=true,height=55mm]{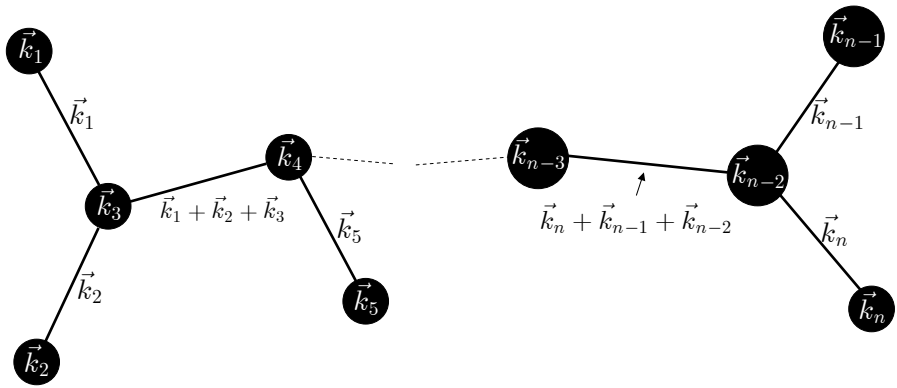}
  \end{center}
  \caption{This figure shows the {\it reduced} tree diagram corresponding to
 one of the leading terms which constitute
  the $n$-point function whose arguments are ${\vec k_1},{\vec
 k_2},\cdots, {\vec k_n}$. The assignment of wavenumbers to the vertices and
 lines is illustrated. }
\label{fig:tree.eps}
\end{figure}

The rules of the reconstruction of the leading term that constitutes
the $n$-point function from a given {\it reduced} tree diagram are 
as follows~\cite{Byrnes:2007tm}. 
First, 
we assign a different wavenumber ${\vec k_i} (1 \le i \le n)$ to each
vertex $\bullet$ of the diagram, where
${\vec k_1},{\vec k_2},\cdots, {\vec k_n}$
are the arguments of the $n$-point function with the constraint
${\vec k_1}+{\vec k_2}+ \cdots + {\vec k_n} = 0$.
Next we assign a wavenumber to each line in the diagram, too.
%If the line is an external one, i.e. if it is attached to one of the end
%points of the diagram, 
%we assign to the line the same wavenumber that is assigned to the end
%point. 
%For an internal line, 
In general, 
removing a line from the diagram yields two respectively connected
sub-diagrams.
Then, one assign to the removed line the sum of the vectors associated with all
vertices in one of the two sub-diagrams. 
We do not care which of two sub-diagrams we choose since only the length of the 
assigned waved number is used in the following discussion. 
An example of the assignment of the wavenumber is given in
Fig.~\ref{fig:tree.eps}. 

After associating the wavenumbers with all lines, 
now we can assign the corresponding factors to the vertices and the
lines.
As for the vertex with $p$ lines attached, assign the factor 
$N_{a_1 a_2 \cdots a_p}$ to it. 
As for the lines, 
assign $A^{ab} P$, where 
the argument of the power spectrum $P$ is set to the length of the 
wavenumber associated with each line. 
 By multiplying all these factors assigned to vertices and lines, and 
summing up all independent diagrams which are not mutually isomorphic,
%taking $n!$ permutations of the argument wavenumbers, 
%$\{ {\vec k_1},{\vec k_2},\cdots, {\vec k_n}\}$,
we obtain a function of $n$ wavenumbers, 
which constitutes the $n$-point function. 
The indices in $N_{a_1 a_2 \cdots a_p}$ assigned to each vertex 
are contracted with the indices of $p$ neighboring lines. Contraction is performed 
between lower and upper indices as usual. 
 
Here, we did not associate a factor $1/p!$ with the vertex 
$N_{a_1 a_2\cdots a_p}$  from the beginning 
for the following reason. 
A vertex with $p$ lines attached has $p$ 
lower indices to be contracted with the upper indices 
in $A^{ab}P$ associated with the $p$ attached lines. 
These $p$ lines are all to be distinguished because they 
are all labelled with different wavenumbers. 
Therefore there are $p!$ ways of contraction between two sets of 
$p$ indices. If we do not distinguish which indices are 
contracted, 
%i.e. if we do not care about the order of lines 
%attached to the same interaction vertex, 
the factor $1/p!$ 
associated with the vertex is canceled. 
%Hence, we did not associate the factor 
%$1/p!$ with the vertex.

\begin{figure}[htbp]
  \begin{center}
    \includegraphics[keepaspectratio=true,height=50mm]{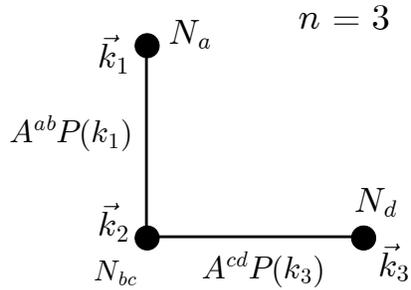}
  \end{center}
  \caption{This diagram represents the leading term of the bi-spectrum of
 primordial curvature perturbations.}
  \label{fig:bis.eps}
\end{figure}

\begin{figure}[htbp]
  \begin{center}
   \includegraphics[keepaspectratio=true,height=50mm]{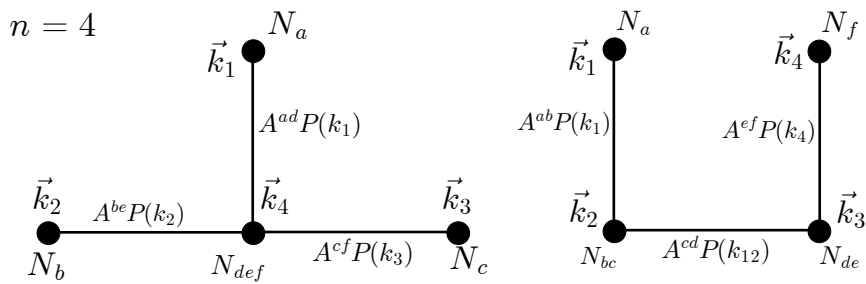}
  \end{center}
  \caption{These diagrams represent the leading order terms of the
 tri-spectrum. These two distinct
 diagrams show different wavenumber dependence. Hence, 
 two parameters are needed to describe the tri-spectrum.}
  \label{fig:tri.eps}
\end{figure}

Finally, 
by taking the sum over all the possible {\it reduced} tree diagrams,
we obtain the {$n$-point function}.
As an illustration,
we show the diagrams for $n=3$ and $4$ 
in Figs.~\ref{fig:bis.eps} and~\ref{fig:tri.eps}, respectively.

It is not a trivial matter whether the functions constructed from 
two {\it reduced} tree diagrams that are not isomorphic to each other
always yield a different functional dependence on the wavenumbers. 
As we explained in the appendix~\ref{proof}, 
if the two {\it reduced} tree diagrams with $n$ vertices are not
mutually isomorphic,
the corresponding functions of $n$ wavenumbers are always
different~\cite{Ota}.
Therefore, 
the number of parameters necessary to determine the $n$-point function of
$\zeta$
is equal to the number of independent {\it reduced} tree diagrams with
$n$
vertices. As an illustration,
we show in Table~\ref{fig:table.eps} the number of free parameters and
the corresponding diagrams for $n=3,4,5$ and $6$.

A similar diagrammatic approach for the higher order correlation functions has been
developed in the context of galaxy correlation or large scale structure.
In Ref.~\cite{Fry:1983cj}, the author has given a numeration of the number of independent tree
diagrams for general $n$, using the generating functions based on the combinatorial analysis~\cite{Riorden:1958}.
Applying this method to our discussion about the higher order correlation functions of the primordial curvature perturbations, we can find the number of independent {\it reduced} tree diagrams
for general $n$, which corresponds to the number of free parameters for general $n$-point functions.

%Such a number of independent {\it reduced} tree diagrams for general $n$
%is not known, and 
%deriving an analytic expression remains an open question in the graph 
%theory of mathematics~\cite{Nagamochi}.

\begin{table}[htbp]
  \begin{center}
    \includegraphics[keepaspectratio=true,height=70mm]{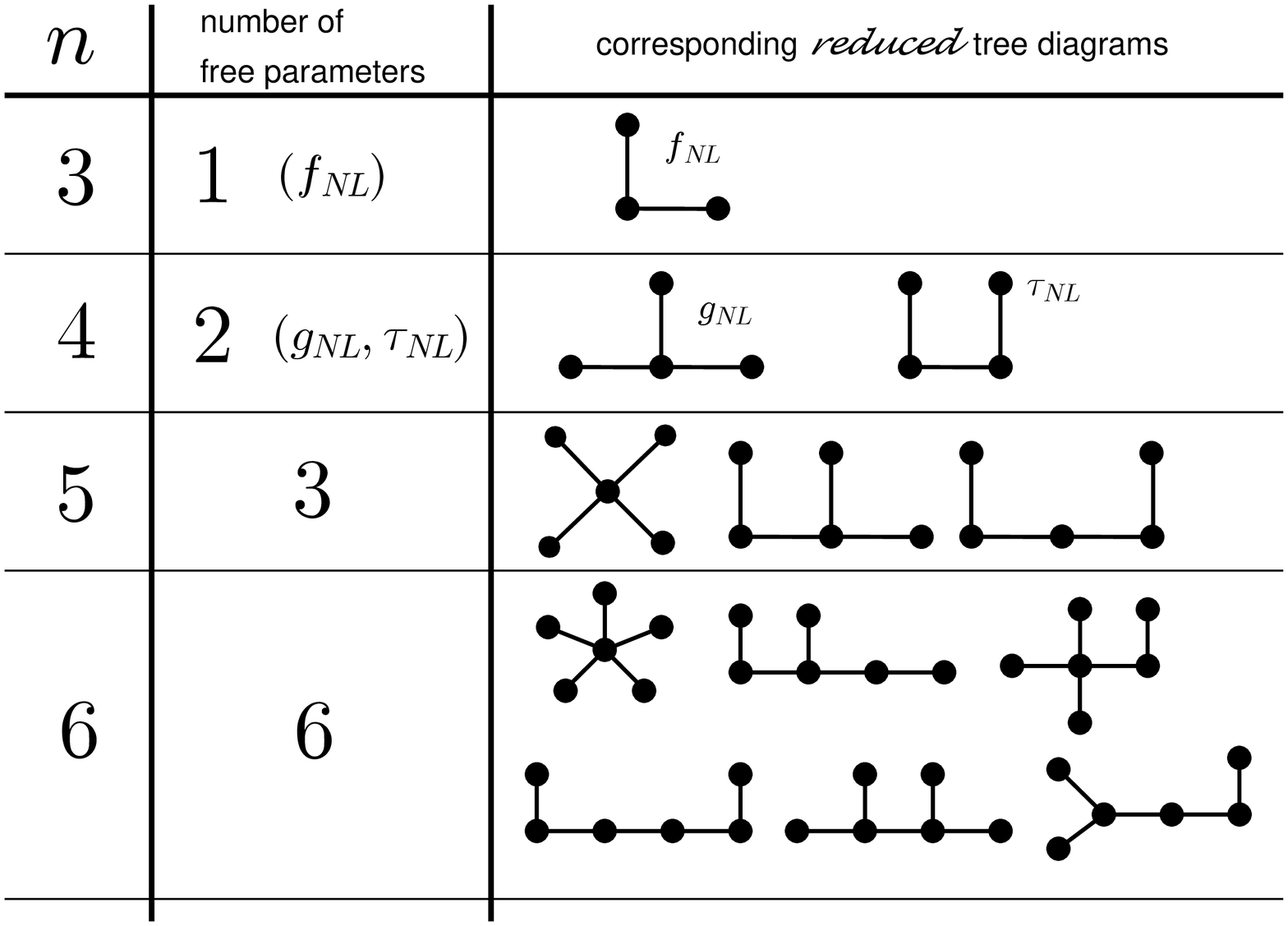}
  \end{center}\vspace*{-5mm}
  \caption{This table shows the number of free parameters and the
 corresponding {\it reduced} tree diagrams for $n$-point correlation functions with
  $n=3,4,5,6$.}
  \label{fig:table.eps}
\end{table}

\section{A new method to compute $n$-point functions}
\label{diagram2} 

In this section,
we will provide an efficient method to compute the non-linearity
parameters to characterize the $n$-point correlation functions. 
%which requires much smaller number of operations proportional to ${\cal N}$. 

\subsection{{\it tree-shaped} diagrams}
\label{treeshaped}
We start with the fact that $\zeta (N_F)$ is independent of the choice
of the time of the initial flat hypersurface $N_*$. 
By choosing $N_*$ to be identical to $N_F$ in (\ref{zeta}), we obtain~\cite{Lyth:2004gb}
\begin{eqnarray}
\zeta (N_F) &=& \sum{1 \over n!} N^F_{a_1a_{2\cdots}a_n}
\delta \varphi^{a_1}_F\delta \varphi^{a_2}_F \cdots \delta
\varphi^{a_n}_F~, 
%\nonumber \\
%&=&\sum{1 \over n!} N^*_{a_1a_{2\cdots}a_n}
%\delta \varphi^{a_1}_*\delta \varphi^{a_2}_* \cdots \delta \varphi^{a_n}_*,
\label{alternativezeta}
\end{eqnarray}
where $\delta \varphi^a_F=\delta \varphi^a(N_F)$ are field perturbations
evaluated {\it on the flat slice} at $N=N_F$.
As we have mentioned in the previous section,
$\zeta(N_F)$ represents the curvature perturbation {\it on the uniform energy density slice}
at $N=N_F$.
Hence the above equation~(\ref{alternativezeta})%KOKO
means that the curvature perturbation $\zeta (N_F)$ is simply caused by 
a time shift between the flat slice and the uniform energy density slice
at the final time $N=N_F$.
%We will use the expression on the first line  in
%Eq.~(\ref{alternativezeta}) rather than that on the second line
%to calculate the non-linearity parameters.
Hence, as shown in the appendix~\ref{Nab},
$N^F_{a_1a_{2\cdots}a_n}$ can be written only by local quantities at $N=N_F$.
$N^F_{a_1a_{2\cdots}a_n}$ are therefore obtained immediately, 
once we specify $\varphi_F^a$. %%changed

What we need to evaluate is $\{\delta \varphi^a_F\}$ as functions 
of $\{\delta \varphi^a_*\}$. 
The evolution equations for $\delta \varphi^a$, which can be obtained by perturbing 
the background equation~(\ref{back1}), are given by
\begin{eqnarray}
{d \over dN}\delta \varphi^a(N) &\!=&\!  P^a_{~b}\delta \varphi^b(N)  + {1 \over 2}Q^a_{(3)bc}(N)\delta \varphi^b(N) \delta \varphi^c(N)+\nonumber\\
&&\cdots +  {1 \over (\ell-1) !}Q^a_{(\ell)b_1b_{2} \cdots b_{\ell-1}}(N)\delta \varphi^{b_1}(N)
\delta \varphi^{b_2}(N) \cdots \delta \varphi^{b_{\ell-1}}(N) + \cdots,
\label{III2}
\end{eqnarray}
where $P^a_{~b}$ and $Q^{a}_{(\ell)b_1b_2\cdots b_{\ell-1}}$ are, respectively, defined by
\begin{eqnarray}
\left.P^{a}_{~b} \equiv {\partial F^a \over \partial
 \varphi^b}\right\vert_{\varphi=\varphizero(N)}~,~
 ~Q^{a}_{(\ell)b_1b_2\cdots b_{\ell-1}}(N) \equiv {\partial^{\ell-1} F^a \over  
   \partial \varphi^{b_1}\partial \varphi^{b_2} \cdots\partial\varphi^{b_{\ell-1}}}\biggr|_{\varphi=\varphizero (N)}~.
\label{PQ}
\end{eqnarray}
%As examples, the explicit forms of $P^a_{~b}$, $Q^a_{(3)bc}$ and $Q^a_{(4)bcd}$ are shown in appendix~\ref{specific}.
For the purpose of the evaluation of the $n$-point function,
it is enough to truncate the expansion on the right hand side in Eq.~(\ref{III2})
at $(n-1)$-th order.
By solving the above equations from $N=N_*$ to
$N_F$ with initial conditions $\delta \varphi^a (N_*)=\delta \varphi^a_*$,
we obtain $\delta \varphi^a_F$ expressed in terms of $\{\delta \varphi^a_*\}$. %%changed 
Due to the non-linear evolution after the horizon crossing,
the distribution of $\{\delta \varphi^a_F\}$ is in general non-Gaussian
even if that of $\{\delta \varphi^a_*\}$ 
is Gaussian. %%changed (moved)

%Since the perturbed equations are non-linear,
%$\delta \varphi^a_F$ depends on $\{\delta \varphi^a_*\}$
%nonlinearly. %%changed (erased)
If we solve the equations (\ref{III2}) %%changed
iteratively,
we can express formally $\delta \varphi^a_F$ as a Taylor expansion
in terms of $\{\delta \varphi^a_*\}$.
Let us denote the $m$-th order terms in the iterative expansion by $\delta \varphim^{a}_F$.
Then $\delta \varphi^a_F$ can be written as
$
\delta \varphi^a_F=\sum_{m=1}^{n-1}~\delta \varphim^{a}_F \label{taylor}
$, 
where we truncate the expansion at the $(n-1)$-th order because higher order terms
are irrelevant to the $n$-point function. By definition, 
$\delta \varphim^{a}_F$  contains 
$m$ Gaussian random variables, $\{\delta \varphi^{a}_*\}$. 
Namely, there is a factor $\delta \varphi^{a_1}_* \cdots \delta
\varphi^{a_m}_*$ in $\delta \varphim^{a}_F$. 
The indices in this factor are to be contracted with 
the interaction vertices $Q^a_{(\ell)b_1 b_2\cdots b_{\ell-1}}$ 
or the Green function $\Lambda^a_{~b}$,  
which obeys 
\begin{eqnarray}
&&{d \over dN}\Lambda^{a}_{~b}(N,N') = P^{a}_{~c}(N)\Lambda^{c}_{~b}(N,N')~, \\
&&{d \over dN'}\Lambda^{a}_{~b}(N,N') = - \Lambda^{a}_{~c}(N,N')P^{c}_{~b}(N')~. 
\end{eqnarray} 
Here again an upper index is contracted with a lower index, as usual. 
All the possible ways of contraction contribute to $\delta \varphim^{a}_F$.

\begin{figure}[tbp]
  \begin{center}
    \includegraphics[keepaspectratio=true,height=75mm]{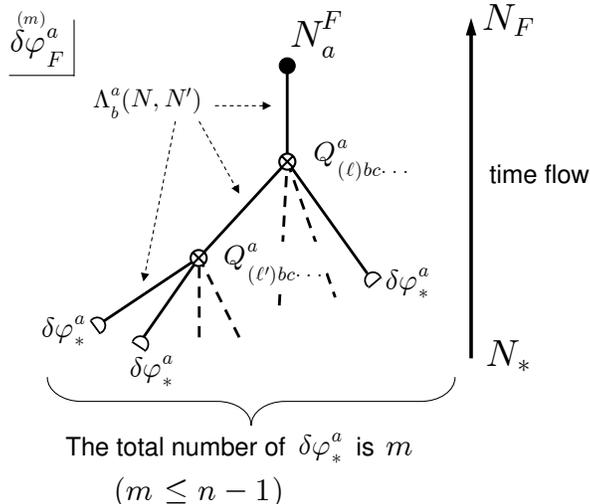}
  \end{center}\vspace*{-10mm}
  \caption{One example of \textit{tree-shaped} diagram, which
 contributes to $\delta \varphim^a_F$. 
%The subscript $(\ell)$ in $Q^a_{(\ell)bc\cdots}$
%  represents the number of the lines which connect the vertex, $\otimes$. To each line segment, the propagator $\Lambda^a_{~b}(N,N')$
%  is assigned. As a rule of drawing the diagram,
 The diagram branches off from the top to the bottom. }%
  \label{fig:ETdia_1.eps}
\end{figure}

We can associate a diagram as presented in Fig.~\ref{fig:ETdia_1.eps}  with
each way of contraction. 
Hereinafter, we refer to such a diagram as a \textit{tree-shaped} diagram,  
to distinguish it from the {\it reduced} tree diagram introduced earlier  
and from %%changed
what is simply called a tree diagram which will be introduced later. 
A \textit{tree-shaped} diagram is drawn obeying the following simple rules.  
We start with a solid circle $\bullet$ 
and attach a line downward to it. 
We attach an interaction vertex $\otimes$ 
to the other end of this line. 
From the vertex, several lines extend downward and they end 
with another interaction vertex $\otimes$ or a half open circle. 
This process is repeated until all the end points are terminated by a half open
circle. The total number of half open circles should be $m$. 

The solid circle $\bullet$ 
corresponds to $N_a^F$, and hence we assign the time $N=N_F$ to it. 
A half open circle corresponds to the initial Gaussian variables
$\delta\varphi^a_*$, and hence the time $N=N_*$ is assigned to it. 
An interaction vertex $\otimes$ corresponds to the factor $Q^b_{(\ell) b_1 \cdots b_{\ell-1}}(N)$,
where $N$ ($N_* \leq N \leq N_F$) is the time assigned to this vertex 
and $\ell$ ($3 \leq \ell \leq m$) is the number of attached lines. 
Here, we did not associate a factor $1/(\ell-1)!$ with 
$Q^b_{(\ell) b_1 \cdots b_{\ell-1}}(N)$.
The reason is the same as before in the case of $N_{a_1 a_2\cdots
a_p}$ discussed in Sec.~\ref{highersec}. 
Here, the half open circles are all supposed to be labelled,
i.e. distinguishable. In this case dropping the factor 
$1/(\ell-1)!$ is exactly compensated by not distinguishing   
the order of lines attached to the same interaction 
vertex.

In this diagram the time flows from the bottom to the top as is indicated in 
Fig.~\ref{fig:ETdia_1.eps}.  
Finally, to each line segment,  
we assign the Green function (propagator) $\Lambda^a_{~b}(N,N')$, where 
$N$ and $N'$ are the time coordinates assigned to the upper and lower
ends of the line, respectively. 
Contracting upper and lower indices between the adjacent objects
(the solid circle, interaction vertices, half open circles and lines) 
and integrating over all the time coordinates assigned to 
the interaction vertices in the whole range of their possible variation, 
we obtain a quantity which constitutes $N_a^F\delta \varphim^{a}_F$. 
Collecting all terms corresponding to different diagrams 
yields total $N_a^F\delta \varphim^{a}_F$.

%This can be easily understood in the following way.  
%A $\ell$-point interaction vertex has $(\ell-1)$ 
%lower indices to be contracted with the upper indices 
%in the propagators associated with the $(\ell-1)$ attached lines extending downward. 
%These propagators are all to be 
%distinguished (even if the shapes of the associated sub-diagrams are  
%the same) because we assumed that the half open circles are all labelled. 
%Therefore there are $(\ell-1)!$ ways of contraction between two sets of 
%$(\ell-1)$ indices. If we do not distinguish which indices are 
%contracted, i.e. if we do not care about the order of lines 
%attached to the same interaction vertex, the factor $1/(\ell-1)!$ 
%associated with the vertex is eliminated. Hence, hereafter we remove the factor 
%$1/(\ell-1)!$ from the interaction vertex $\otimes$. 

\subsection{$n$-point correlation functions}
\begin{figure}[ptb]
  \begin{center}
    \includegraphics[keepaspectratio=true,height=40mm]{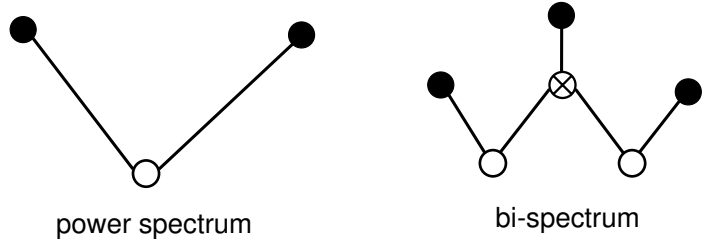}
  \end{center}\vspace*{-10mm}
  \caption{Tree diagrams corresponding to the power- and bi-spectra.}
  \label{fig:ETdia_2.eps}
\end{figure}

Instead of $\zeta(N_F)$, 
we first compute $n$-point functions of $\zeta^{(\rm lin)}_F$
defined by 
\begin{eqnarray}
\zeta^{(\rm lin)}_F =N_a^F \delta\varphi^a_F,
\end{eqnarray}
which is the linear truncation of the Taylor expansion of $\zeta (N_F)$ 
in terms of $\delta \varphi^a_F$.
The $n$-point function of $\zeta^{(\rm lin)}_F$ is given 
by the sum of all the possible connected tree diagrams obtained 
by contracting all the half open circles $\{\delta \varphi_*^a\}$ in pair 
from a product of $n$ \textit{tree-shaped} diagrams. (See Fig.~\ref{fig:ETdia_2.eps}.)
Contraction between the half open circles 
within the same \textit{tree-shaped}
diagram can be neglected because it produces a loop. 
For the same reason, there is not more than one contraction between 
any pair of \textit{tree-shaped} diagrams.  
Let us represent this contraction between a pair of half open circles, 
by a full open circle $\circ$, to which $A^{ab}$ defined in Eq.~(\ref{aij})
is assigned. 
We refer to the diagram obtained by this contraction simply as a tree diagram. 
The leading terms of the $n$-point function of $\zeta^{(\rm
lin)}_F$ are ${\cal O}(\delta^{2(n-1)})$, and hence 
the tree diagram should have $(n-1)$ open circles $\circ$.
As any pair of \textit{tree-shaped} diagrams does not have more than one
contraction between them, all the half open circles belonging to a 
single \textit{tree-shaped} diagram 
are contracted with different \textit{tree-shaped} diagrams. 
Since all the \textit{tree-shaped} diagrams are labelled with 
a different wavenumber $\vec k_i$, %%changed 
the assumption that all the  half open circles are 
distinguishable holds. This contraction process 
does not produce any further statistical weight, and hence 
all the tree diagrams have the same weight of unity.

In addition to the linear term $\zeta^{(\rm lin)}_F$,
$\zeta (N_F)$ contains terms non-linear in $\delta \varphi^a_F$, 
which also contribute to the $n$-point functions of $\zeta (N_F)$.
There is one-to-one correspondence between 
the non-linear terms in $\zeta (N_F)$ and the interaction 
vertex $\otimes$ that is directly connected to $\bullet$ by a line 
without any intervening vertices $\circ$ or $\otimes$. %%changed
Hence, we can take into account this non-linear contribution simply by 
replacing the interaction vertices directly connected to $\bullet$ as 
\begin{eqnarray}
N^F_a \Lambda^a_{~c} (N_F,N) Q^c_{(\ell) b_1 \cdots b_{\ell-1}} (N)
 \longrightarrow && 
N^F_a \Lambda^a_{~c} (N_F,N) \hat Q^c_{(\ell) b_1 \cdots b_{\ell-1}} (N)\cr
 && \equiv  N^F_a \Lambda^a_{~c} (N_F,N) Q^c_{(\ell) b_1 \cdots
 b_{\ell-1}} (N)+N^F_{b_1 \cdots b_{\ell-1}} \delta~(N-(N_F-\varepsilon)),
\label{nonlinrule}
\end{eqnarray} 
where $\varepsilon$ is an infinitesimally small number. 
By this prescription,
we can obtain the non-linearity parameters only from the tree diagrams.

Now we are ready to show that our method to evaluate the $n$-point functions can be reduced to 
the problem of solving the ordinary differential equations for vector variables
that have only a single index $a$ ($1\leq a \leq 2{\cal N})$, which is 
the main result of this paper. 
Let us consider one tree diagram which constitutes the $n$-point
function. 
We focus on one of sub-diagrams obtained by removing one 
vertex $\otimes$
or $\circ$ from a tree diagram,  
which we denote by $\Gamma_a$ or $\tilde{\Gamma}^a$.
If the line which was attached to the removed object is pointing
downward (upward), the vector has a lower (an upper) index. 
Suppose that the object attached to the other end of 
this line is an interaction vertex $\otimes$ 
with which $Q^a_{(\ell)b_1 \cdots b_{\ell-1}}(N)$ associates. %%changed
Let us consider the case of a vector $\tilde{\Gamma}^a$ with an upper index. 
Notice that the other lines connected to this vertex are also
similar sub-diagrams which consist of smaller number of vertices
than that we are focusing on. 
We denote the product of the vectors associated with all these
sub-diagrams by a tensor $M^{c_1\cdots c_{\ell-1}}$. 
Those vectors appearing in $M^{c_1 \cdots c_{\ell-1}}$ are already known
by the induction assumption. 
Then, $\tilde \Gamma^a$ can be defined recursively as 
\begin{eqnarray}
\tilde{\Gamma}^{a}(N)=\int_{N_*}^N dN'~\Lambda^{a}_{~b}(N,N') Q^b_{(\ell) c_1 \cdots c_{\ell-1}}(N') M^{c_1 \cdots c_{\ell-1}} (N'). \label{veceq}
\end{eqnarray}
From this equation,
we find that $\tilde{\Gamma}^{a} (N)$ satisfies 
\begin{eqnarray}
\frac{d}{dN} \tilde{\Gamma}^{a}(N)
  =P^{a}_{~b}(N)
 \tilde{\Gamma}^b(N)+Q^{a}_{(\ell) c_1 \cdots c_{\ell-1}} (N) M^{c_1 \cdots c_{\ell-1}}
 (N). 
\label{veceqdif}
\end{eqnarray}
The boundary conditions for $\tilde{\Gamma}^{a}(N)$ are set by
$\tilde{\Gamma}^{a}(N_*)=0$ at $N=N_*$, and hence the above equation is
to be solved in the forward direction in time. 

In the case of the vector with an upper index $\tilde\Gamma^a(N)$, the neighboring object can be
$\circ$ instead of $\otimes$. In this case the initial conditions are 
given by $\tilde\Gamma^a(N_*)=A^{ab}\Gamma_b(N_*)$, where
$\Gamma_b(N_*)$ is the vector corresponding to the sub-diagram with 
the neighboring vertex $\circ$ being removed. The equation to solve is
simply the homogeneous one given by 
\begin{eqnarray}
{d \over dN} \Gamma_a(N) = - \Gamma_b(N)P^b_{~a}(N). 
\end{eqnarray}

Similarly, for a vector $\Gamma_a$ with the neighboring object being $\otimes$, we have
\begin{eqnarray}
\Gamma_a(N) = \int^{N_F}_{N} dN'~\Lambda^{c_1}_{~a}(N',N)
\hat Q^b_{(\ell)c_1 \cdots c_{\ell-1}}(N')M_b^{~c_2 \cdots c_{\ell-1}}(N')~,
\end{eqnarray}
and this vector obeys 
\begin{eqnarray}
{d \over dN} \Gamma_a(N) = - \Gamma_b(N)P^b_{~a}(N) - \hat Q^b_{(k)a c_{2} \cdots c_{\ell-1}}(N)M_b^{~c_2 \cdots c_{\ell-1}}(N)~.
\end{eqnarray}
The boundary conditions for $\Gamma_a(N)$ are set by
$\Gamma_a(N_F) = 0$ at $N=N_F$, or equivalently 
$\Gamma_a (N_F - \epsilon) = N_{a c_2 \cdots c_{\ell-1}} $ $
M^{c_2 \cdots c_{\ell-1}}(N_F)$
taking into account the $\delta$-function term in the definition of 
$\hat{Q}^c_{(\ell)b_1\cdots b_{\ell-1}}$ in Eq.~(\ref{nonlinrule}) 
as boundary conditions. 
In this manner, the effect of the non-linear terms in $\zeta(N_F)$ 
in (\ref{alternativezeta}) can be absorbed by the boundary conditions 
in general. %%changed
%The boundary conditions for ${\Gamma}_{a}(N)$ are set by
%{\bf
%${\Gamma}_{a}(N_F)=0$ at $N=N_F$},
The equation is solved backward in time. 
There is another case in which the neighboring object is $\bullet$. 
This simplest case can be also handled in a similar manner. 
%However, %%changed
We defer its explanation to the succeeding section, 
where we exhibit some more explicit formulas. %%changed (change line)
In Table~\ref{fig:ETdia_7.eps}, 
we summarize the notation of the vector quantities 
which will be used below, showing the correspondence 
to the \textit{tree-shaped} diagram. 

\begin{table}[htbp]
  \begin{center}
    \includegraphics[keepaspectratio=true,height=80mm]{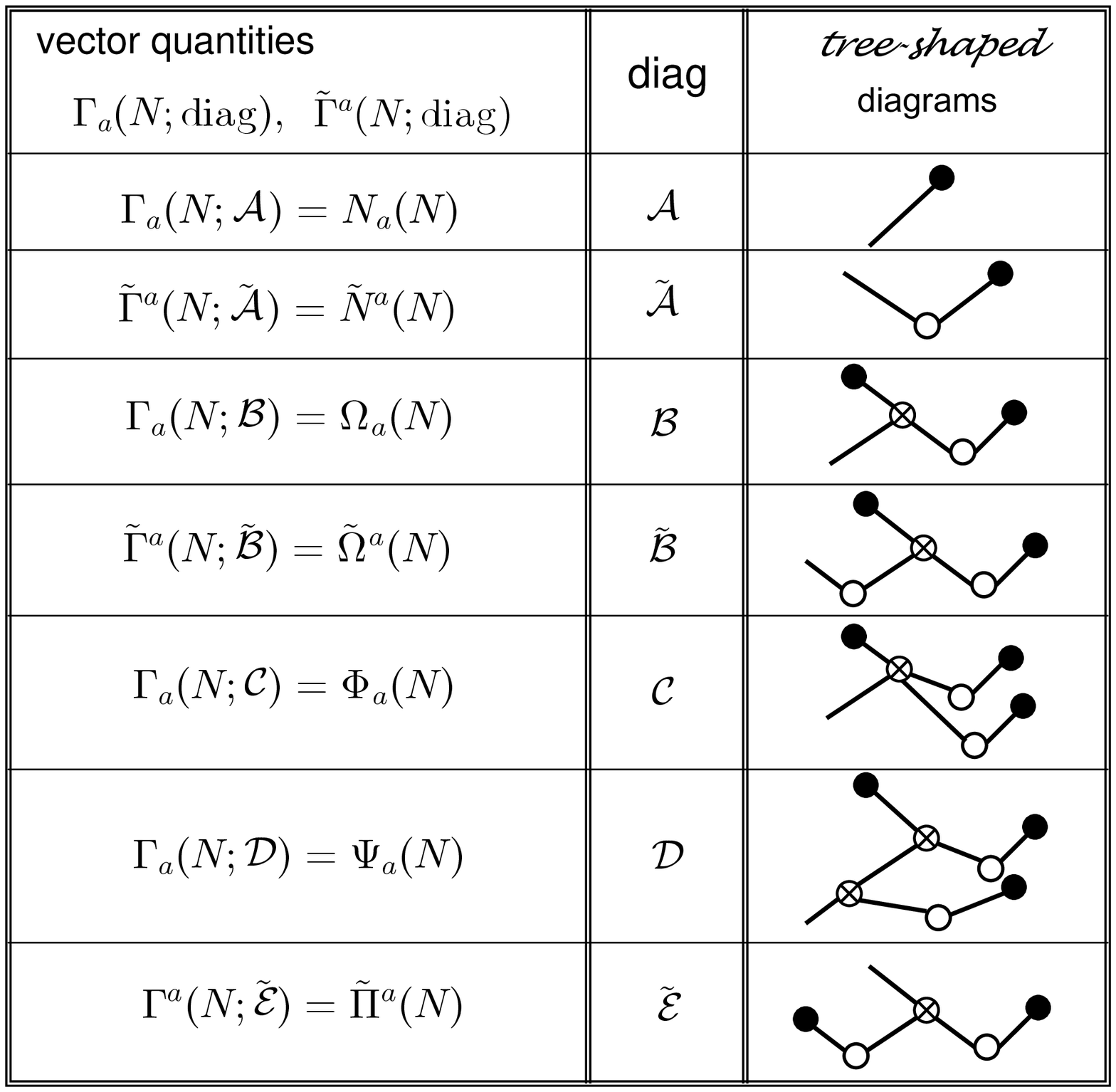}
  \end{center}\vspace*{-5mm}
  \caption{This table shows the summary of the correspondence between
 the vector quantities used in this paper. }
  \label{fig:ETdia_7.eps}
\end{table}

To obtain an expression for the $n$-point function 
written in terms of such
vectors, we arbitrarily choose one vertex $\otimes$ or $\circ$ 
from a tree diagram at the beginning. 
Suppose that the chosen vertex is an interaction vertex $\otimes$ with which 
$Q^a_{(\ell)b_1 \cdots b_{\ell-1}}(N)$ associates.
After preparing all the necessary vectors, $\Gamma_a$ and
$\tilde\Gamma^{b_i}_{(i)}$ $(1 \leq i \leq \ell-1)$, which  
correspond to the sub-diagrams obtained when this vertex is removed, %%changed
we can immediately 
write down the contribution to the $n$-point function from this tree
diagram as  
\begin{eqnarray}
\int^{N_F}_{N_*} dN~\Gamma_{a}(N)Q^a_{(k)b_1 \cdots
 b_{\ell-1}}(N)\prod_{i=1}^{\ell-1} \tilde\Gamma^{b_i}_{(i)}(N)~. 
\label{finalsum}
\end{eqnarray}
If the vertex which we initially focused on is $\circ$, 
with which $A^{ab}$ associates, we do not need the final integration over $N$. 
We denote the vectors that correspond to the sub-diagrams obtained by
removing this $\circ$ by $\Gamma^{(1)}_{a}$ and $\Gamma^{(2)}_{b}$. 
Then, we compute 
\begin{eqnarray}
A^{ab} \Gamma^{(1)}_a(N_*) \Gamma^{(2)}_b(N_*), 
\label{finalsum2}
\end{eqnarray}
instead of the expression~(\ref{finalsum}).
In this diagrammatic method, 
the final expression for the spectrum in appearance depends on which 
vertex we chose at the beginning, but, of course, all different 
looking expressions are equivalent. 
Practically, it is more efficient to choose a vertex near the
center so as to reduce the number of necessary vectors, 
although the definition of the center of a diagram is 
not so clear in many cases.  

%{\bf (I think arrows are not necessary any more.)
%To clarify which formula we should use, it might be convenient 
%to attach an arrow to the propagators, although it is redundant. 
%The arrow is pointing toward the focused 
%vertex $\otimes$ or $circ$ which we chose at the beginning. 
%Then, we can solve the vectors one by one starting with $\bullet$ along the
%arrow.  
%Since the time flows from bottom to top, 
%when the arrow at the tip of the sub-diagram is pointing upward (downward),
%then the corresponding vector is solved by integrating in the forward
%(backward) direction in time.
%}

\subsection{Relation to the {\it reduced} tree diagrams and
  statistical weight}

As we mentioned in Sec.~\ref{highersec}, the {\it reduced} tree diagram
is useful to classify the wavenumber dependence of the $n$-point functions,
while the ({\it full}) tree diagram is a powerful tool for explicit
computation of the $n$-point functions. 
We show that 
the {\it reduced} tree diagram 
introduced in Sec.~\ref{highersec} 
is actually a simplified version of the tree diagram. 
It will be manifest that the solid circles attached to the ends of 
diagrams have the same meaning in both diagrams. 
In the {\it reduced} tree diagram the internal lines represent 
power spectrum of the initial Gaussian random field $\{\delta\varphi_*^a\}$, 
which is expressed by an open circle $\circ$ in the tree diagram.
Hence, each line in the {\it reduced} tree diagram corresponds 
to a line with an open circle $\circ$ in the tree diagram.
The sub-structure described by the interaction vertices $\otimes$ in the tree diagram 
is completely abbreviated in the {\it reduced} tree diagram. 
Hence, there is a degeneracy such that different tree diagrams
contribute to the same {\it reduced} tree diagram. 
As explained in Sec.~\ref{highersec} and proven in
appendix~\ref{proof}, the wavenumber dependence of the $n$-point
functions is classified by the topology of the {\it reduced} tree
diagram. This means that plural tree diagrams can give the contribution 
to $n$-point function with the same wavenumber dependence. 

\begin{figure}[htbp]
  \begin{center}
    \includegraphics[keepaspectratio=true,height=80mm]{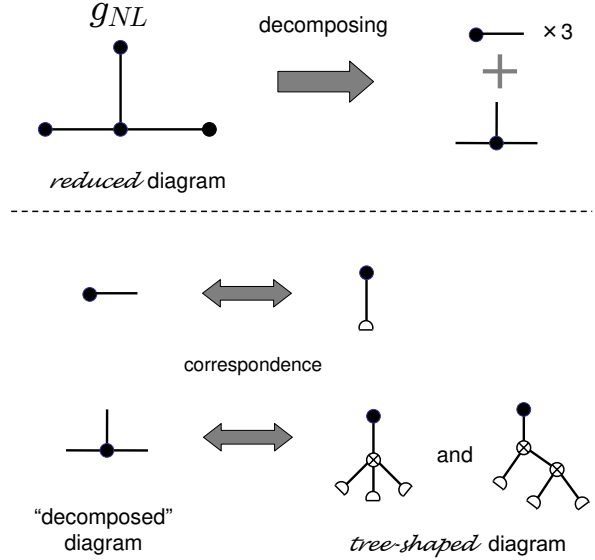}
  \end{center}\vspace*{-5mm}
  \caption{This figure shows the relation between the
 \textit{tree-shaped} diagram and the {\it reduced} diagram for the
 non-linearity parameter $g_{NL}$. The upper one shows how to decompose
 the {\it reduced} diagram, and the lower one shows the correspondence
  between the "decomposed" diagrams and the \textit{tree-shaped} diagrams.}
  \label{fig:ETdia_4.eps}
\end{figure}

In Fig.~\ref{fig:ETdia_4.eps}, 
as an example, we show the diagrams corresponding to the
tri-spectrum coefficient $g_{NL}$.
We can decompose the top-left {\it reduced} tree diagram 
into 4 sub-diagrams by cutting all lines off.
These sub-diagrams are counter parts of the \textit{tree-shaped} diagrams.
The lower part of this figure explains correspondence 
between these sub-diagrams and 
the \textit{tree-shaped} diagrams. 
There are two \textit{tree-shaped} diagrams with four half open circles 
as is explicitly shown in this figure.
Hence, we find that the formula for $g_{NL}$ is 
composed of two different terms. 

When we consider the statistical weight of the diagram, this
correspondence between the {\it reduced} and {\it full} tree diagrams is important. 
The starting point is the fact that the statistical weight of each tree
diagram is unity when each end point $\bullet$ is labelled 
by the assigned momentum. 
Therefore counting the statistical weight by writing down all 
different tree diagrams is straightforward. 
However, the non-linearity parameters are defined based on 
the {\it reduced} tree diagram. 
The most of the patterns which occur as a result of permutation of 
the momenta assigned to the end points $\bullet$ is taken care already in 
the definition of the non-linearity parameters. (See Eqs.~(\ref{higher2}) 
and (\ref{II-14}).) However, here we should notice that  
some of the half open circles in 
the {\it tree-shaped} diagrams can be distinguishable, while 
the sub-diagrams obtained from the {\it reduced} tree diagram as
mentioned above do not distinguish their legs at all. 
Therefore when there are several distinguishable patterns to assign 
the labels to the half open circles in a \textit{tree-shaped} diagram, 
the term containing such a \textit{tree-shaped} diagram has a factor 
corresponding to the number of patterns. 

As an example, we again consider the case of $g_{NL}$. 
The \textit{tree-shaped} diagram that has two 
three-point interaction vertices
shown in Fig.~\ref{fig:ETdia_4.eps} 
has three distinguishable patterns in assigning the labels $\{1,2,3\}$ to
the three half open circles. 
This means that 
we need to add the corresponding factor $3$ to the contribution
containing this \textit{tree-shaped} diagram. (See the
expression in Eq.~(\ref{maingf}) below.) 

We want to emphasize that 
the above formulation simplifies the computation of higher order
correlation functions a lot. %%changed  
In this formulation, we have only to solve vector quantities with 
only one index. 
Therefore our computation scheme requires the number of operations 
proportional to ${\cal N}$ in computing the non-linearity parameters in
$n$-point function. 
If we performed a naive straightforward calculation, in which 
the derivatives of the $e$-folding number $N$ are computed by 
using the finite difference method numerically, 
the required number of operations is proportional to ${\cal N}^n$.   
If we naively performed perturbative expansion, in which 
we connect the interaction vertices by propagators $\Lambda^a_{~b}(N,N')$ and 
perform integration over the time coordinates of the interaction vertices, 
the necessary number of operations would be even larger. 
When the number of inserted interaction vertices is $m$, 
naively we need 
$O({\rm several}~ \times~ {\rm (number~of~time~steps)}^m)$ operations 
to compute the contribution of the single diagram.
On the other hand, in our formulation we need only 
$O({\rm several}~ \times~ {\rm (number~of~time~steps)}~\times (m+{\rm a~few}))$ operations.
Therefore our scheme is particularly useful for 
the computation of higher order correlation functions in the inflation models 
with a large number of field components. %%changed ${\cal N}$. 

\section{Examples}
\label{examplesec}
In this section 
we apply our formalism to the computation of the power-, bi-, 
tri- and {\it quad}-spectrum to demonstrate the efficiency of our method. 

\subsection{Power spectrum}
Let us first consider the power spectrum.
There is only one tree diagram that contributes to the power spectrum,
which is shown on the left hand side in Fig.~\ref{fig:ETdia_2.eps}.
Following the prescription given in the previous section, we focus on a
unique vertex $\circ$.

\begin{figure}[ptb]
  \begin{center}
    \includegraphics[keepaspectratio=true,height=30mm]{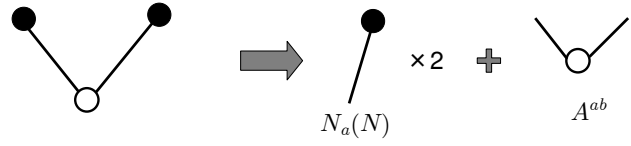}
  \end{center}\vspace*{-5mm}
  \caption{This figure shows how we decompose the tree diagram for
 the power spectrum, when we focus on the vertex $\circ$. }%
  \label{fig:ETpow.eps}
\end{figure}

Then, we can decompose this tree diagram into this open circle, with
which $A^{ab}$ associates, and two identical sub-diagrams shown on the right
hand side in Fig.~\ref{fig:ETpow.eps}. 
Corresponding to this simplest sub-diagram, 
we introduce a vector $N_a(N)$, 
whose explanation was deferred in the preceding
section.  
This vector is defined by the equation 
%\begin{eqnarray}
%N_a (N) \equiv \Lambda^b_{~a} (N_F,N) N^F_b.
%\end{eqnarray}
%This vector obeys the following homogeneous equation;
\begin{eqnarray}
\frac{d}{dN}N_a(N)=-P^b_{~a}(N) N_b (N), \label{eq:n}
\end{eqnarray}
with the boundary conditions $N_a(N_F)=N_a^F$.  
The vector $N_a(N)$ represents the derivatives of the $e$-folding number 
with respect to $\varphi^a$ evaluated at $\varphi^a=\varphizero^a(N)$. 
Using this vector, the power spectrum of $\zeta (N_F)$ is expressed as~\cite{Yokoyama:2007dw}
\begin{eqnarray}
\frac{P_\zeta}{P}=A^{ab}N_a(N_*)N_b(N_*) \equiv W_*, \label{power1}
\end{eqnarray}
where $P_\zeta$ and $P$ are those which have already appeared in
Eq.~(\ref{higher1}). 

\subsection{Bi-spectrum} 
\begin{figure}[tbp]
  \begin{center}
    \includegraphics[keepaspectratio=true,height=40mm]{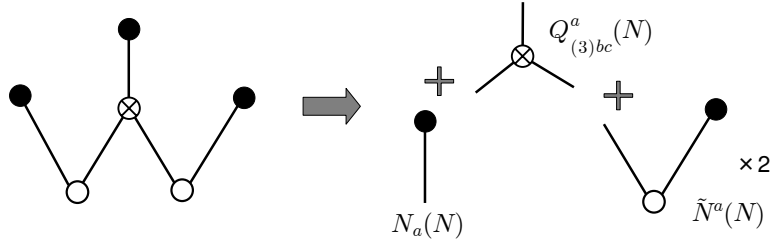}
  \end{center}\vspace*{-5mm}
  \caption{This figure shows how we decompose the tree diagram for 
 the bi-spectrum, when we focus on the interaction vertex $\otimes$. }
  \label{fig:ETdia_3.eps}
\end{figure}

Just like the power spectrum,
there is only one tree diagram that contributes to the bi-spectrum, which
is presented on the right hand side in Fig.~\ref{fig:ETdia_2.eps}. 
Let us focus on the interaction vertex $\otimes$ to which $Q^a_{(3)bc}(N)$ is
assigned, and decompose the diagram 
into the chosen vertex $\otimes$, a sub-diagram denoted by $N_a(N)$ and
the two same sub-diagrams denoted by $\tilde{N}^a(N)$ as illustrated in
Fig.~\ref{fig:ETdia_3.eps}. 

The sub-diagram denoted by $\tilde{N}^a(N)$ is
reduced to that denoted by ${N}_a(N)$ if we remove one open circle $\circ$.
Hence, following the general rule explained in the preceding section, 
the new vector $\tilde{N}^a(N)$ is obtained by integrating
%\begin{eqnarray}
%\tilde{N}^a(N)  \equiv \Lambda^{a}_{~c}(N,N_*)A^{cb}N_{b}(N_*)~.
%\label{tildeN}\\
%\end{eqnarray}
%This vector obeys the following homogeneous equation;
\begin{eqnarray}
{d \over dN}\tilde{N}^{a}(N) = P^{a}_{~b}(N) \tilde{N}^{b}(N)~,  
\label{difftn} 
\end{eqnarray}
from $N=N_*$
with the initial conditions $\tilde{N}^a(N_*) = A^{ab}N_b(N_*)$. 
Applying the general formula~(\ref{finalsum}) supplemented by~(\ref{nonlinrule}),
we recover the result previously obtained in
Ref.~\cite{Yokoyama:2007dw},
\begin{eqnarray}
{6 \over 5}f_{NL} = W_*^{-2}\left[N_{ab}^F\tilde{N}^a(N_F)\tilde{N}^b(N_F) 
+ \int^{N_F}_{N_*}dN N_a(N)Q^{a}_{(3)bc}(N)\tilde{N}^b(N)\tilde{N}^c(N)\right]~.
\label{mainf}
\end{eqnarray}

\subsection{Tri-spectrum}  
  
\begin{figure}[htbp]
  \begin{center}
    \includegraphics[keepaspectratio=true,height=60mm]{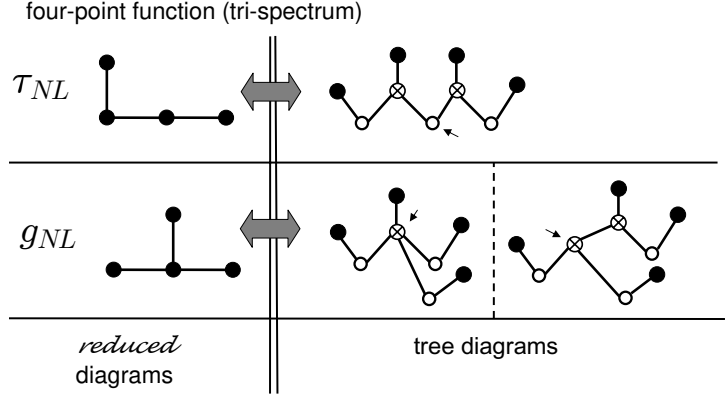}
  \end{center}\vspace*{-5mm}
  \caption{The diagrams for the tri-spectrum. As we have mentioned in the previous
 section~\ref{highersec}, for the tri-spectrum due to the difference of
 the scale dependence we need the two parameters $\tau_{NL}$ and
 $g_{NL}$, which can be also distinguished by using the {\it reduced}
 diagram. Using the tree diagram $g_{NL}$ is decomposed to two
 diagrams as shown in this figure. The arrows indicate the focused vertex in each diagram.}
  \label{fig:ETdia_8.eps}
\end{figure}

As we mentioned in the previous subsection~\ref{highersec},
we need two parameters, $\tau_{NL}$ and $g_{NL}$, for the tri-spectrum.
The tree diagrams for the tri-spectrum were shown in Fig.~\ref{fig:ETdia_8.eps}.
From this figure, 
we find that $g_{NL}$ consists of two tree diagrams.

We choose a focused vertex in each diagram 
as indicated by arrows in Fig.~\ref{fig:ETdia_8.eps}.
Following the prescription explained in the preceding section, 
we have 
\begin{eqnarray}
&&\tau_{NL} = 
W_*^{-3}\left[ A^{ab}\Omega_a(N_*) \Omega_b(N_*) \right],
\label{maintf} \\
&&g_{NL} =
{25 \over 54} W_*^{-3}\Biggl\{
N_{abc}^F \tilde{N}^a(N_F)\tilde{N}^b(N_F)\tilde{N}^c(N_F) + 
\int^{N_F}_{N_*}dN N_{a}(N)Q^{a}_{(4)bcd}(N)\tilde{N}^b(N)\tilde{N}^c(N)\tilde{N}^d(N)
~\nonumber\\
&& \qquad\qquad\qquad\qquad\qquad
+3\int^{N_F}_{N_*}dN\Omega_a(N)Q^a_{(3)bc}(N)\tilde{N}^b(N)\tilde{N}^c(N)\Biggl\}~.\label{maingf}
\end{eqnarray}
where $\Omega_a (N)$ is a new vector obtained by solving 
%defined by
%\begin{eqnarray}
%\Omega_a(N) \equiv N_{bc}^F \tilde{N}^b(N_F) \Lambda^c_{~a}(N_F,N)
%+ \int^{N_F}_{N}dN' N_b(N')Q^{b}_{(3)cd}(N')\Lambda^{c}_{~a}(N',N)\tilde{N}^d(N')~.\label{Omega}
%\end{eqnarray}
%This vector obeys the following inhomogeneous equation;
\begin{eqnarray}
{d \over dN} \Omega_a(N) =  - \Omega_b(N)P^b_{~a}(N)-
 N_b(N)Q^b_{(3)ac}(N)\tilde{N}^c(N)~,\label{diffom}
\end{eqnarray}
backward in time from $N=N_F$ 
with the boundary conditions $\Omega_a(N_F) = N_{ab}^F \tilde{N}^b(N_F)$. 
The first (second) line on the right hand side of $g_{\rm NL}$
represents the contribution of the left (right) tree diagram 
corresponding to $g_{NL}$ in Fig.~\ref{fig:ETdia_8.eps}.
In our formulation, 
it is enough to solve differential equations for
three vectors $N_a,~{\tilde N}^a,~\Omega_a$ to compute 
$\tau_{\rm NL}$ and $g_{\rm NL}$.

\subsection{{\it quad}-spectrum}

\begin{figure}[htbp]
  \begin{center}
    \includegraphics[keepaspectratio=true,height=85mm]{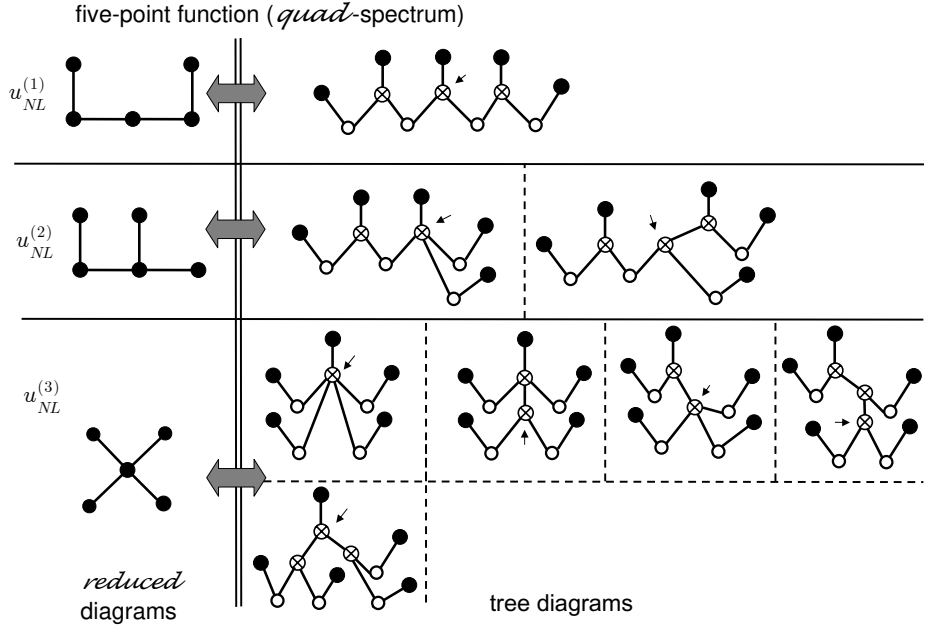}
  \end{center}\vspace*{-5mm}
  \caption{This shows the correspondence between {\it reduced} diagram and
 the tree diagram for the fifth-order spectrum
  as an example. The arrows indicate the focused vertex in each diagram.}
  \label{fig:ETdia_6.eps}
\end{figure}

In order to demonstrate the efficiency of our formulation, we show
the explicit formula for the {\it quad}-spectrum.
We also show the correspondence between the {\it reduced} diagram and the tree
diagram for the fifth-order spectrum (five-point correlation function)
in Fig.~\ref{fig:ETdia_6.eps}.
 %%changed (no line change)
Using the formula, we obtain an expression for the {\it quad}-spectrum 
as
\begin{eqnarray}
\langle \zeta_{{\bf k}_1} \zeta_{{\bf k}_2} \zeta_{{\bf k}_3}
 \zeta_{{\bf k}_4} \zeta_{{\bf k}_5}
  \rangle_c 
&=&  u^{(1)}_{NL}\left( P_\zeta (k_1)
     P_\zeta (k_{12})P_\zeta (k_{45})P_\zeta (k_5) +
						    59~{\rm perms.}
						   \right) \nonumber\\
&&  + u^{(2)}_{NL}\left( P_\zeta (k_1)
     P_\zeta (k_{12})P_\zeta (k_{4})P_\zeta (k_5) +
						    59~{\rm perms.}
						   \right) \nonumber\\
&&+ u^{(3)}_{NL}\left( P_\zeta (k_1)
     P_\zeta (k_{2})P_\zeta (k_{3})P_\zeta (k_4) +
						    4~{\rm perms.}
						   \right), 
\end{eqnarray}
with 
\begin{eqnarray}
&&u^{(1)}_{NL} = 
 W_*^{-4}\Biggl\{
\int^{N_F}_{N_*}dN N_{a}(N)\hat
Q^{a}_{(3)bc}(N)\tilde{\Omega}^b(N)\tilde{\Omega}^c(N)\Biggr\}~, 
\cr
&&u^{(2)}_{NL} = 
 W_*^{-4}
\int^{N_F}_{N_*}dN 
\Biggl\{
N_{a}(N)
 \hat Q^{a}_{(4)bcd}(N)\tilde{\Omega}^b(N)\tilde{N}^c(N)\tilde{N}^d(N)
+3\Omega_{a}(N)
 \hat Q^{a}_{(3)bcd}(N)\tilde{\Omega}^b(N)\tilde{N}^c(N)\Biggr\}~,\cr
&&u^{(3)}_{NL} = 
 W_*^{-4}
\int^{N_F}_{N_*}dN 
\Biggl\{
N_{a}(N)
 \hat Q^{a}_{(5)bcde}(N)\tilde{N}^b(N)\tilde{N}^c(N)\tilde{N}^d(N)\tilde{N}^e(N)
+6\Phi_{a}(N)
 \hat Q^{a}_{(3)bc}(N)\tilde{N}^b(N)\tilde{N}^c(N)\cr
&&\qquad\qquad\quad\quad\quad\quad\quad\quad\quad 
+4\Omega_{a}(N)
 \hat Q^{a}_{(4)bcd}(N)\tilde{N}^b(N)\tilde{N}^c(N)\tilde{N}^d(N)
 +12\Psi_{a}(N)
 \hat Q^{a}_{(3)bc}(N)\tilde{N}^b(N)\tilde{N}^c(N)\cr
&&\qquad\qquad\quad\quad\quad\quad\quad\quad\quad\quad\quad 
+3 N_a(N)\hat{Q}^a_{(3)bc}(N)\tilde{\Pi}^b(N)\tilde{\Pi}^c(N) 
\Biggr\}~.
\end{eqnarray}
Here we introduced new vectors defined by the equations  
\begin{eqnarray}
&&\frac{d}{dN} \tilde\Omega^a (N)=P^a_{~b} (N) \tilde\Omega^b (N), \\
&&\frac{d}{dN} \Phi_a (N)=-P^b_{~a} (N) \Phi_b (N)-  N_b (N) Q^b_{(4) cd
 a}(N) {\tilde N}^c (N) {\tilde N}^d (N),\label{diffphi} \\
&&\frac{d}{dN} \Psi_a (N)=-P^b_{~a} (N) \Psi_b (N)-\Omega_b (N) Q^b_{(3)
 ca} (N) {\tilde N}^c (N),\label{diffpsi} \\
 &&\frac{d}{dN} \tilde{\Pi}^a(N) = P^a_{~b} (N) \tilde{\Pi}^b(N) + Q^a_{(3)bc}\tilde{N}^b(N)\tilde{N}^c(N),\label{diffpi} \\
\end{eqnarray}
with the boundary conditions $\tilde\Omega^a (N_*)=A^{ab}\Omega_b
(N_*)$, 
$\Phi_a(N_F) = N_{abc}^F \tilde{N}^b(N_F)\tilde{N}^c(N_F)$,
$\Psi_a(N_F) = 0$ and $\tilde{\Pi}^a(N_*) = 0$.

\section{Discussion and conclusion}
\label{sum}

The primordial non-Gaussianity has been focused on by many authors as a
new probe of the inflation dynamics.
The deviation from the Gaussian statistics affects not only the
bi-spectrum of the primordial curvature perturbations 
but also the higher order correlation functions.
In general, 
to describe the higher order correlation functions, 
we need more parameters and more complicated calculations.
Instead of resorting to the direct calculations, 
we developed a diagrammatic method, 
which is useful in 
counting the number of necessary non-linearity parameters and 
computing the higher order correlation functions for non-Gaussianity of
local type. 
We showed that the number of parameters to describe the $n$-point
correlation function is equal to the number of {\it reduced}
tree diagrams with $n$ vertices that are not isomorphic to each other.
We also found that in the calculation of general $n$-point correlation
function 
we have only to solve the vector quantities which follow the same linear 
perturbation equation for the background field or it's 
dual~\cite{Yokoyama:2007dw}, but with a source term and different
boundary conditions.
Our formalism requires the number of operations proportional to 
${\cal N}$ even for higher order correlation functions, 
in contrast to the naive expectation $\propto {\cal N}^n$, where 
${\cal N}$ is the number of components of the inflaton field. 
It will be clear that our formulation is particularly powerful for the
inflation models with many components of scalar field,
including the models in which
the slow-roll conditions are violated after the horizon crossing time. 

\begin{figure}[htbp]
  \begin{center}
    \includegraphics[keepaspectratio=true,height=30mm]{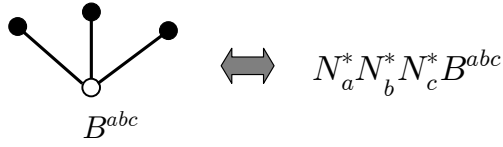}
  \end{center}
  \caption{This diagram corresponds to the leading order correction due
 to the
 non-Gaussianity of the initial perturbation $\delta \varphi^a_*$ to
  the bi-spectrum.}
  \label{fig:ETdia_5.eps}
\end{figure}

In this paper, we assumed that the distribution of 
initial perturbations of the field $\{\delta \varphi_*^a\}$ is Gaussian.
As a results, in the diagram the number of lines connected to the
open circle, which corresponds to the contraction of $\delta \varphi_*^a$, is two. 
When we need to consider the effects of non-Gaussianity of $\delta \varphi_*^a$,
we can easily extend our formalism by adding open circles with
appropriate numbers of the attached lines. 
For example, 
it has been well known that the leading effect 
of non-Gaussianity in $\delta \varphi_*^a$ 
affects the three-point correlation function as~\cite{Byrnes:2006vq,Seery:2006vu,Seery:2006js}
\begin{eqnarray}
&&\langle \zeta_{{\bf k}_1}\zeta_{{\bf k}_2}\zeta_{{\bf k}_3}\rangle
\equiv \left[N_a^*N_b^*N_c^*B^{abc}(k_1,k_2,k_3) + {6 \over
	5}f_{NL}^{G}\left(P_\zeta(k_1)P_\zeta(k_2) + 2~{\rm perms.}
		    \right)
\right]\delta^{(3)}({\bf k}_1+{\bf k}_2+{\bf k}_3)~,
\label{fullbispectrum} \\
&&\langle \delta \varphi^a_* \delta \varphi^b_* \delta \varphi^c_*\rangle
\equiv B^{abc}(k_1,k_2,k_3)\delta^{(3)}({\bf k}_1+{\bf k}_2+{\bf k}_3)~,
\label{fullbispectrum2} 
\end{eqnarray}
where ${6 \over 5}f_{NL}^{G}$ denotes the non-linearity parameter given
by Eq.~(\ref{mainf}), which has been obtained under
the assumption that
$\delta \varphi_*^a$ is Gaussian.
In our diagrammatic method, the first term on the right hand side of Eq.~(\ref{fullbispectrum})
can be described by the diagram presented in
Fig.~\ref{fig:ETdia_5.eps}. 
For this open circle $\circ$ with three legs we assign the factor
$B^{abc}$ defined in (\ref{fullbispectrum2}). 
Generalization of taking into account the higher order correlators is 
straightforward. Application of our formulas to some explicit models
will be reported in the forthcoming paper. 

%Our formulation can be also applied to the models in which the curvature
%perturbation is generated at the end of 
%inflation~\cite{Bernardeau:2002jf,Lyth:2005qk,Salem:2005nd,Alabidi:2006wa,Alabidi:2006hg,Matsuda:2008hx,Sasaki:2008uc,Naruko:2008sq}.
%For such cases, it has been reported that large $f_{NL}$ can be generated.
%Hence, it will be worth calculating the higher order non-linearity 
%parameters such as $\tau_{NL}$ and $g_{NL}$ and obtaining relations between 
%these parameters in this context~\cite{Ichikawa:2008iq}.
%We plan to show the results of these calculations in a forthcoming paper~\cite{forthcoming}. 

\begin{acknowledgments}

The authors would like to thank Professor Katsuhiro Ohta
at Keio University for providing the proof in appendix~\ref{proof}.
The authors would also like to thank Professor Hiroshi Nagamochi 
at Kyoto University for notifying us an important mathematical fact.
SY is grateful to Takahiko Matsubara and Tsutomu Takeuchi for useful comments. 
SY is supported in part by Grant-in-Aid for Scientific Research
on Priority Areas No. 467 ``Probing the Dark Energy through an
Extremely Wide and Deep Survey with Subaru Telescope''.
He also acknowledges the support from the Grand-in-Aid for the Global COE Program
``Quest for Fundamental Principles in the Universe: from Particles to the Solar
 System and the Cosmos '' from the
Ministry of Education, Culture, Sports, Science and Technology (MEXT) of
Japan. 
TT is supported 
by Monbukagakusho Grant-in-Aid
for Scientific Research Nos.~17340075 and~19540285. %kiban(B)
He also acknowledges the
support from the Grant-in-Aid for the Global COE Program ``The Next
Generation of Physics, Spun from Universality and Emergence'' from MEXT of
Japan. 

\end{acknowledgments}

\appendix
\section{Proof of subsection~\ref{highersec}}
\label{proof}

We give a proof of the statement that  
the functions obtained by applying the rules in Sec.~\ref{highersec}
to two tree diagrams with $n$ vertices that are not isomorphic 
to each other show different wavenumber dependence~\cite{Ota}.
To prove it,
it is enough to show that we can uniquely reconstruct the tree 
diagram with $n$ vertices from a given function $f({\vec k_1},\cdots,{\vec k_n})$,
which guarantees one-to-one correspondence between a diagram
and a function $f({\vec k_1},\cdots,{\vec k_n})$.
%To avoid the inessential complexity,
%we assume that only one field contributes to $\zeta$ and
%$N_{\phi \cdots \phi}$ are all unity.
Here each wavenumber ${\vec k_i}$ is assigned to each vertex. 
By construction, the function 
$f$ should be a product of $(n-1)$ power spectra, $P$, whose arguments are
the length of the sum of several wavenumbers taken from ${\vec k}_i (1 \leq i \leq n)$.

Let us focus on one arbitrary vertex of the would-be reconstructed diagram.
We refer to this vertex as $V_m$ 
and the vector attached to this vertex as ${\vec k}_m$. 
We eliminate ${\vec k}_m$ from the arguments of $P$ by using the relation ${\vec k}_1+\cdots +{\vec k}_n=0$.
Then, the wavenumber assigned to a vertex connected to $V_m$ by a line 
must appear in $f$ only once. This is because such a wavenumber 
appears only in $P$ corresponding to the line that connects this vertex to $V_m$. 
By finding all such wavenumbers, 
%which we denote by ${\vec k}_{m_1},\cdots,{\vec k}_{m_n}$,
we recognize all the vertices that are connected to the vertex $V_m$. 
% is connected to 
% $\{V_{m_i}\}$ with $1 \leq  i \leq  n$. 
By doing the same thing for each vertex, 
we completely recognize how all the vertices are mutually connected. 
Obviously, this fixes the shape of the diagram uniquely.

\section{Explicit formulas for derivatives of $N^F$}
\label{Nab}
As mentioned in Sec.~\ref{treeshaped},
$N_{a_1a_{2\cdots}a_n}^{F}$ defined in 
Eq.~(\ref{alternativezeta}) can be written in terms of local quantities evaluated at
$N=N_F$.
Here, as examples, we explicitly evaluate the coefficients 
$N_{a}^{F},N_{ab}^{F}$ and $N_{abc}^{F}$.
Taking the hypersurface at $N=N_F$ to be a uniform Hubble one, 
which is equal to the uniform density slicing on super-horizon scales, 
we have the equation,  
\begin{eqnarray}
H \left( \varphi^a (N_F+\zeta(N_F)) \right)=H \bigl( \varphizero^a (N_F) \bigr)~. \label{exp1}
\end{eqnarray}
The Hubble parameter $H$ is given by Eq.~(\ref{friedmann}).
In our previous paper~\cite{Yokoyama:2007dw}, 
solving Eq.~(\ref{exp1}) with respect to $\zeta(N_F)$ up to the second order, 
we have obtained
\begin{eqnarray}
&&N_{a}^{F}=-\frac{H_a(\varphi)}{H_b(\varphi)
 F^b(\varphi)}\biggr|_{\varphi=\varphizero (N_F)}~, \label{exp4} \\
&&N_{ab}^{F}=-\frac{U_{ab}(\varphi)}{H_c
 (\varphi)F^c(\varphi)}\biggr|_{\varphi=\varphizero (N_F)}~,
 \label{exp5}
\end{eqnarray}
where 
\begin{eqnarray}
U_{ab}=H_{ab}+2\left( H_{c} P^{c}_{~a}+ F^c H_{ca}
				\right)N^F_{b}+\left( F^c H_{cd} F^d
+ H_c P^{c}_{~d}F^d \right)N_a^FN_b^F~, \label{exp3}
\end{eqnarray}
with $H_a \equiv \partial H/ \partial \varphi^a,~H_{ab} \equiv
\partial^2 H/ \partial \varphi^a \partial \varphi^b $. 
%The explicit forms of $H_{a}$ and $H_{ab}$ are 
%shown in appendix \ref{specific}.
Solving Eq.~(\ref{exp1}) up to the third order, we also obtain
\begin{eqnarray}
N_{abc}^{F} = -{W_{abc}(\varphi) \over H_d(\varphi)
 F^d(\varphi)}\biggr|_{\varphi=\varphizero (N_F)}~, \label{exp6}
\end{eqnarray}
where
\begin{eqnarray}
W_{abc} &\!=\!& H_{abc} 
+\left[H_d\left(Q^d_{(3)ef}F^e+P^d_{~e}P^e_{~f}\right)F^f + H_{def}F^dF^eF^f
+ 3 F^dH_{de}P^e_{~f}F^f\right]N_a^FN_b^FN_c^F~\nonumber\\
&&
+3
\left[2F^dH_{de}P^e_{~a}+\left(H_{ade}F^d+H_{ad}P^{d}_{~e}\right)F^e
+H_d\left(Q^d_{(3)ea}F^e + P^d_{~e}P^e_{~a}\right)\right]N^F_bN^F_c, 
~\nonumber\\
&&
+{3}\left(2H_{ad}P^d_{~b}+F^dH_{dab}+H_dQ^d_{(3)ab}\right)N_c^F~\nonumber\\
&&
+3\left(F^dH_{de}F^e + H_d P^d_{~e}F^e\right)N^F_aN^F_{bc}+3\left(F^dH_{da}+H_dP^d_{~a}\right)N^F_{bc}
\end{eqnarray}
with $H_{abc} \equiv {\partial^3 H / \partial \varphi^a \partial \varphi^b \partial \varphi^c}$.
Note that $U_{ab}$ and $W_{abc}$ are symmetric with respect to the indices.
%The explicit forms of $H_a$, $H_{ab}$ and $H_{abc}$ are given in appendix~\ref{specific}.
As we mentioned earlier, here we define the phase space variables as
$\varphi^I_1 = \phi^I$ and $\varphi^I_2 = \dot{\phi}^I$.

\end{document}